\documentclass[pre,showpacs,twocolumn,preprintnumbers,amsmath,amssymb]{revtex4}

\usepackage{graphicx}
\usepackage{dcolumn}
\usepackage{bm}
\newcommand{\comment}[1]{}

\begin{document}

\preprint{APS/123-QED}

\title{Mechanical rejuvenation and over-aging in the soft glassy rheology model}

\author{Mya Warren}
\email{mya@phas.ubc.ca}
\author{J{\"o}rg Rottler}%

\affiliation{Department of Physics and Astronomy, The University of
British Columbia, 6224 Agricultural Road, Vancouver, BC, V6T 1Z1,
Canada}

\date{\today}

\begin{abstract}
Mechanical rejuvenation and over-aging of glasses is investigated
through stochastic simulations of the soft glassy rheology (SGR)
model. Strain- and stress-controlled deformation cycles for a wide
range of loading conditions are analyzed and compared to molecular
dynamics simulations of a model polymer glass. Results indicate that
deformation causes predominantly rejuvenation, whereas over-aging
occurs only at very low temperature, small strains, and for high
initial energy states. Although the creep compliance in the SGR model
exhibits full aging independent of applied load, large stresses in the
nonlinear creep regime cause configurational changes leading to
rejuvenation of the relaxation time spectrum probed after a stress
cycle. During recovery, however, the rejuvenated state rapidly returns
to the original aging trajectory due to collective relaxations of the
internal strain.
\end{abstract}

\maketitle

\section{Introduction}
The mechanical properties of glassy materials continuously evolve due
to slow, non-equilibrium dynamics, a phenomenon called physical aging
\cite{angell_JAP,Struik,Hutchinson}. As a consequence, the response of
glasses to an applied load depends not only on measurement time, but
also on the wait time $t_w$ that has elapsed since the glass was
formed. In general, increasing wait time makes glasses less compliant
and increases their resistance to plastic flow
\cite{Struik,Hutchinson,Cloitre_PRL85,Bonn_PRL89,Gacougnolle_Poly43,McKenna_JPhys15,Lequeux_PRL89,Lequeux_Farad123,Lacks_PRL93,Utz_PRL84,Lyulin_PRL99,Rottler_PRL95,Warren_PRE07}. For
glasses formed through a rapid quench from the liquid state, the effects of aging take a particularly simple form: response functions such as the creep compliance obey a self similar scaling with the wait time and
depend only on the ratio of $t/t_w^\mu$. The aging exponent $\mu$ has
been found to be approximately unity for a wide class of structural
glasses moderately below the glass transition temperature
\cite{Struik, Schweizer07}.

However, large mechanical deformation and plastic yielding can modify
the aging dynamics. A primary example of this phenomenon is the
observed reduction in the aging exponent obtained through creep
measurements of glassy polymers at large stress. Since the relaxation
times of the deformed glass resemble those of a younger glass, it was
hypothesized that the glass had been ``rejuvenated'' by the stress
\cite{Struik}. Experiments on a wide variety of materials including
polymer glasses
\cite{McKenna_Poly31,Gacougnolle_Poly43,Ediger_JCP128}, colloidal
glasses \cite{Cloitre_PRL85,Bonn_PRL89}, and even the cytoskeleton of
the cell \cite{Weitz_NatureMat4} have shown a similar decrease in
relaxation times under high loads. However, the interpretation of
these results in terms of rejuvenation remains controversial. Detailed
experiments by McKenna have shown that the time to equilibration of
mechanically rejuvenated glasses remains essentially unchanged by
the application of external load \cite{McKenna_JPhys15}, suggesting that stress
does not in fact change the extent of aging. Other studies indicate
that the apparent enhancement of particle mobility eventually
disappears after unloading \cite{Gacougnolle_Poly43,Ediger_JCP128},
and that the configurational states of mechanically rejuvenated
glasses are distinct from the states visited by aging in the absence
of load \cite{Lacks_PRL93,Lacks_PRL96,Lyulin_PRL99}.

Signatures of over-aging due to deformation have recently been
observed as well. For instance, polymer glasses subject to a stress
relaxation experiment well below the glass transition temperature
exhibit rapid densification for certain strains
\cite{McKenna_PES37}. Molecular simulations of simple structural
glasses show that a small amplitude strain cycle at zero temperature
can decrease the inherent structure energy
\cite{Lacks_PRL93}. Also, detailed experiments by Lequeux and co-workers show over-aging in systems of dense colloids, which are
effectively athermal glasses. They observe changes to the entire
spectrum of relaxation times \cite{Lequeux_PRL89,Lequeux_Farad123}:
after a period of small amplitude oscillatory shear, the glass appears
rejuvenated over small timescales, and over-aged over longer
timescales.

The emerging picture of the phenomenology of glasses under load
presents some interesting questions that seem to defy a simple
explanation in terms of the rejuvenation hypothesis. Firstly, under
what loading conditions do over-aging and rejuvenation occur? So far,
over-aging has been seen only under very specific circumstances: low
temperature, small strains, and strain-controlled loading
conditions. Rejuvenation, on the other hand, has been observed much
more generally in deformed samples, but has been studied most
extensively at high temperatures (only moderately below the glass
transition temperature) and constant stress conditions. Another important
question is presented by the work of McKenna
\cite{McKenna_JPhys15}. What is the nature of the states created by
loading, and how do we reconcile the rejuvenated relaxation spectrum
with the fact that the time to equilibration is
unchanged by the stress?

A comprehensive molecular model that describes aging and deformation
in glasses is not available to date. However, phenomenological energy
landscape approaches such as the soft glassy rheology (SGR) model
\cite{Sollich_PRL78} have been able to capture many aspects of the
rheology of glasses, and have successfully been used to interpret
over-aging in strain-controlled experiments at low temperature
\cite{Lacks_PRL96,Lequeux_Farad123}. In this study, we will use the SGR model to systematically explore the loading phase diagram in order to better understand the generality and the implications of mechanical rejuvenation and over-aging, as well as the relationship between configurational and dynamical changes. We perform
stochastic simulations of the SGR model over a wide range of
experimental conditions and compare the results to molecular dynamics
simulations of a model polymer glass in selected cases.

\section{Models}
\subsection{Soft Glassy Rheology Model}

It has been found experimentally and through molecular simulations
that the structural relaxation events which result in aging and
plastic deformation involve the cooperative motion of groups of
approximately 10-30 particles \cite{Weeks_JCM15,Vollmayr_EPL76,Weitz_Science318}. The
premise of the SGR model is that each of these mesoscopic rearranging
domains can be described by a single fictive particle in a rough
free-energy landscape. This particle performs thermally or
mechanically activated hops between locally harmonic ``traps'' in the
landscape. The density of states of the traps is
\begin{equation}
	\rho(E) = \frac{1}{x_g}\exp(-E/x_g),\qquad E\ge 0
	\label{eq:rho}
\end{equation}
where $x_g$ is the glass transition temperature, and $E$ is the depth
of a trap. At low temperature, many traps will be very long
lived. The master equation governing the dynamics of the fictive
particles is \cite{Sollich_PRL78}
\begin{eqnarray}\nonumber
	\dot{P}(E,l,t)&=&-\dot{\gamma}\frac{\partial P}{\partial
	l}-P\Gamma_0
	\exp\left[-\left(E-\frac{kl^2}{2}\right)/x\right]\\&+&\Gamma(t)\rho(E)\delta(l).\label{master-eq}
\end{eqnarray}
$P(E,l,t)$ is the occupation probability of a state with energy $E$
and local strain $l$ at time $t$. The energy barrier for particle
hopping is reduced by the local strain energy, $E_b=E-kl^2/2$ where
$k$ is the stiffness of the well. All of the barriers have a common
energy of zero, so the energy of the fictive particle in the trap is
$-E_b$. On the right hand side of eq.~(\ref{master-eq}), the first
term describes the elastic motion of the fictive particles in their wells under
a strain rate of $\dot{\gamma}$. In the absence of particle hopping,
this term simply increases the local strain variable of the
particles. The second term in eq.~(\ref{master-eq}) describes activated
hopping out of the wells, which can be viewed as a local plastic yield
event; $\Gamma_0$ is the attempt rate and $x$ is a ``noise
temperature'' \cite{Sollich_PRL78}. The noise temperature is generally
higher than the thermodynamic temperature, as it incorporates the
effect of non-equilibrium fluctuations in the aging glass. The third
term describes the transition to the new state after hopping, whose
energy is chosen randomly from the density of states $\rho(E)$ and is
initially unstrained; $\Gamma(t)$ is the total hopping rate at time
$t$. The macroscopic stress in this formulation is
\begin{equation}
	\sigma=k\langle l \rangle.
	\label{eq:stress}
\end{equation}

We solved the master equation (\ref{master-eq}) numerically using
Monte Carlo simulations of ten thousand particles under various
thermo-mechanical conditions. In this work, $x_g$, $\Gamma_0$, and $k$
are chosen to be one, which has the effect of setting the units of
energy, time, and strain. Note that a strain of one in these units is
the yield strain of an average particle. An initial configuration was
created from the liquid state at $x_{l}>1$ from the corresponding
Boltzmann distribution and was then quenched instantaneously to the
glass phase at $x<1$. In the absence of load, the system falls out of
equilibrium and exhibits full aging, i.e. an aging exponent $\mu=1$
\cite{Bouchaud_JPA}.

For the strain-controlled loading protocol, each fictive particle is
treated independently. The strain variable is increased at constant
rate, and at each time step the particles $i$ hop with probability
$P_i=\Gamma_i dt$ where $\Gamma_i=\Gamma_0
\exp\left[-\left(E_i-kl_i^2/2\right)/x\right]$ is the hopping rate of
the particle. The stress is computed from Eq.~(\ref{eq:stress}), where
the average is taken over all 10,000 particles. The stress-controlled
loading protocol is somewhat more complicated because of the implicit
relationship between the master equation and the stress. In this case,
we use the kinetic Monte Carlo method \cite{Bortz_KMC} to evolve an
ensemble of particles. A time step is chosen from a Poisson
distribution with the global rate $\Gamma=\sum \Gamma_i$, and the particle that will
hop is chosen with a probability proportional to each particle's
individual hopping rate. The relaxed particle hops into a zero strain
state, and the stress it was holding is redistributed uniformly among
the particles in the system, i.e. the microscopic strains are
increased evenly until Eq.~(\ref{eq:stress}) is satisfied again. If
the stress is applied quickly, the strain energy may exceed the
barrier for some elements. These are relaxed instantaneously and the
procedure is repeated until a stable configuration is found.

\subsection{Molecular Dynamics}
For qualitative comparison with a structural glass, we perform
molecular dynamics (MD) simulations on a ``bead-spring'' polymer model
\cite{Kremer_Grest} that has been studied extensively for its
glass-forming properties. In this model, beads interact via a
non-specific van der Waals potential (Lennard-Jones), and bonds are
modeled as a stiff spring (FENE) that prevents chain crossing. The
reference length-scale is $a$, the diameter of the bead; the energy
scale, $u_0$, is determined by the strength of the van der Waals
potential, and the time scale is $\tau_{LJ}=\sqrt{ma^2/u_0}$, where
$m$ is the mass of a bead. The pressure and stress are therefore
measured in units of $u_0/a^3$.

In these simulations, we consider 855 chains of 100 beads each in an
originally cubic simulation volume with periodic boundary
conditions. The polymers are first equilibrated at a melt temperature
of $1.2u_0/k_B$, and then quenched into the glassy state by decreasing
the temperature at constant rate to below the glass transition
temperature $T_g \approx 0.35u_0/k_B$
\cite{Rottler_PRE64}. Deformation experiments are then performed on the glass at constant temperature by either applying a uniaxial load, or by imposing volume-conserving, uniaxial deformation at constant strain rate. This polymer model has been used
frequently in studies of deformation of polymer glasses
\cite{Schweizer_MM08}. In particular, it demonstrates mechanical
rejuvenation during nonlinear creep under large loads
\cite{Warren_PRE07}. Except at very large strains where polymer
entanglements become important, we expect these results to be
pertinent for most glassy materials.

\section{Results}
\begin{figure}[tbp]
\begin{centering}
\includegraphics[width=8cm]{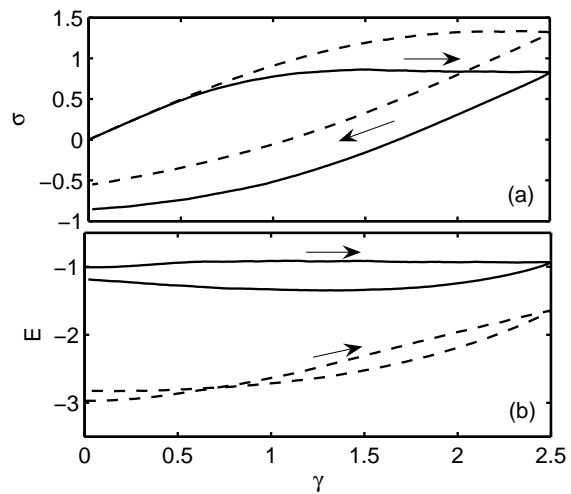}
	\caption{(a) Stress vs.~strain and (b) Energy
	vs.~strain at zero noise temperature. The solid line is for an
	initial configuration at a liquid temperature of $x_l=2$, and
	the dashed line corresponds to $x_l=1.5$.}
	\label{fig:lacks}
	\end{centering}
\end{figure}

We begin exploring the loading phase diagram in the limit of zero
noise temperature and zero strain rate, where over-aging has most
commonly been observed. Starting configurations are created from
different melt temperatures $x_l$. At $x=0$, the samples are strained
at constant rate to a maximum strain $\gamma_{max}$ and then returned
to zero strain at the same rate. The stress and the mean energy are
plotted versus strain in Fig.~\ref{fig:lacks} for two different
initial melt temperatures. Initially, the stress is linear in strain
as the system responds elastically, and then becomes constant after
yield. Note that the yield stress is higher for the state that was
quenched from the lower melt temperature. Also, the energy of the
higher $x_l$ configuration is lowered by the strain cycle
(over-aging), whereas the lower $x_l$ configuration has a higher
energy after the same cycle (rejuvenation). These results are
qualitatively similar to molecular simulation results of a binary
Lennard-Jones glass at zero temperature reported in
ref.~\cite{Lacks_PRL93} and appear to be generic to rough energy
landscapes with many metastable states \cite{Lacks_PRL96}.

\begin{figure}[tbp]
\begin{centering}
\includegraphics[width=8cm]{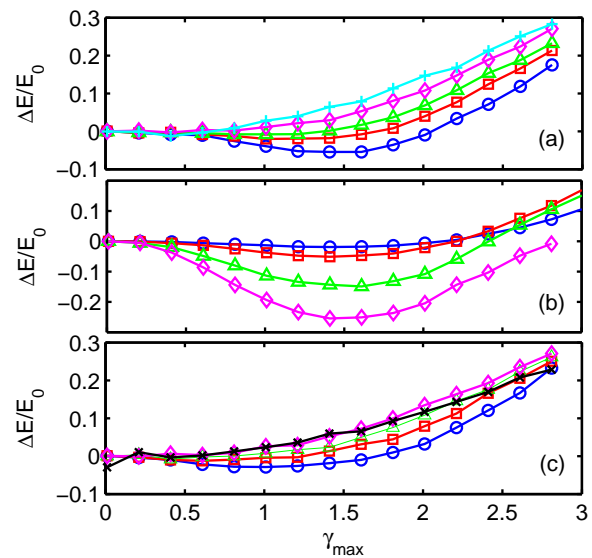}
	\caption{(Color online) Relative difference in energy versus maximum
	strain after a single strain cycle of amplitude
	$\gamma_{max}$. (a) Noise temperature $x=$ 0.1, 0.2, 0.3,
	0.5, and 0.8 from bottom to top, $x_l=2$ and $\dot{\gamma}=$
	0.25 for each sample. (b) $x_l = $ 1.25, 1.5, 2.5, and 5
	from top to bottom, $x=$ 0.001 and $\dot{\gamma}=$ 0.25 for each sample. (c)
	$\dot{\gamma} = $ 1 (circle), 0.5 (square), 0.25 (triangle),
	0.1 (diamond), and 0.01 (cross), $x=$ 0.5 and $x_l=$ 2.}
	\label{fig:strain}
	\end{centering}
\end{figure}

In Fig.~\ref{fig:strain}, we explore in detail the effects of the
noise temperature $x$, the initial configuration temperature $x_l$,
and the strain rate $\dot{\gamma}$ in the strain controlled protocol
described above. Here we treat $x$ as a free adjustable parameter,
although in the SGR model it is envisioned to be related to the
dissipated energy of yielding elements.  For each
thermo-mechanical history, we compare the final energy after the
strain cycle $E_f$ to the energy $E_0$ of the same sample if it had
not been strained, but simply aged at constant temperature for an
equivalent amount of time. The relative energy difference $\Delta E /
E_0$ is positive for rejuvenation, and negative for over-aging.

In Fig.~\ref{fig:strain}(a) the energy difference $\Delta E / E_0$ is
plotted for samples with the same $x_l$ and $\dot{\gamma}$, but
various noise temperatures $x$. We see that over-aging does indeed
occur only at low noise temperatures, which may be why it has mostly
been observed in low temperature simulations and in colloids. The
small $x$ curves show a transition from over-aging at low strains to
rejuvenation at high strains. At very high strains (not shown) the
glass yields, and all of the samples are maximally rejuvenated. As the
noise temperature is increased, the magnitude of over-aging and
the maximum strain where it occurs both decrease. For $x>1$, the SGR
model is in equilibrium and the effects of over-aging and rejuvenation
disappear except for weak transients.

Figure \ref{fig:strain}(b) shows the energy difference at very low $x$
for various initial states $x_l$. The amount of over-aging decreases
as the initial energy is decreased, and asymptotically approaches zero
in the case of a quench from exactly $x_l=x_g$. Note that the strain
at which over-aging is maximized does not seem to depend on the
initial configuration, but always occurs at $\gamma_{max}\approx
1.4$. Alternatively, at higher temperatures where rejuvenation is
predominant, we find that the amount of rejuvenation increases with
melt temperature $x_l$.

Finally, the mechanical rate also plays a role in whether the energy
will be reduced or increased by the strain cycle. Figure
\ref{fig:strain}(c) shows relative energy changes for a sample
with noise temperature $x=0.5$ for various strain rates. The amount of
rejuvenation decreases with increasing strain rate for moderate
strains, but the curves eventually cross as the strain approaches
yield. This is due to the fact that the yield stress (or strain)
increases with rate. At the highest strain rates, over-aging appears
possible even at high temperatures. Note, however, that high strain
rates are not treated entirely realistically in this model. In the SGR
model, an element yields instantaneously when strained to greater than
the yield strain. However, in real solids, very fast strain rates lead
to large affine displacements but relatively few plastic events as
these require a finite time to relax.

\begin{figure}[tbp]
\begin{centering}
\includegraphics[width=8cm]{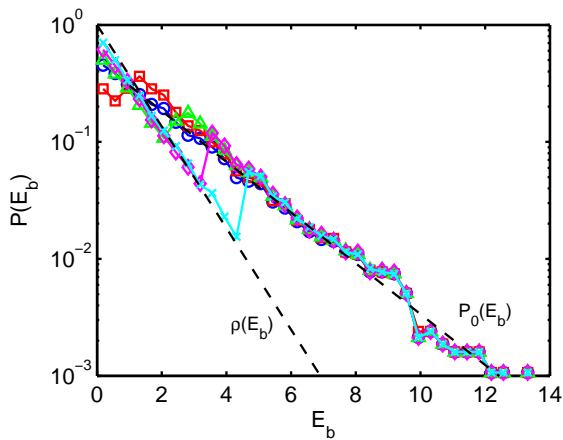}
	\caption{(Color online) Distribution of energy barriers after a zero
	temperature strain cycle. $\gamma_{max}$ goes from zero (dark
	blue circles) to 3 (light blue $\times$'s). The initial energy
	distribution $P_0(E)$ and the landscape density of states
	$\rho(E)$ are shown as dashed lines.}
	\label{fig:P_E}
	\end{centering}
\end{figure}

With the SGR model, further intuition can be gained by investigating
the effect of mechanical deformation on the population of the traps
directly. Figure \ref{fig:P_E} shows the distribution of energy
barriers for various $\gamma_{max}$ after a zero noise temperature
strain cycle. The strain cycle causes particles in the
lowest energy states to hop into new states chosen from the landscape
energy distribution $\rho(E)$. Over-aging then occurs when the states
being relaxed are of lower energy than the mean energy of the states
they hop into.  At zero noise temperature, this predicts that maximum
over-aging occurs at a strain amplitude $\gamma_{max}$ where
$k\gamma_{max}^2/2=\int_0^\infty dE \rho(E) E = 1$, or
$\gamma_{max}=\sqrt{2}$.  This can be seen clearly in Fig.~\ref{fig:strain}(b) for all values of $x_l$. The strain amplitude at
maximum over-aging thus gives a measure of the mean energy of the
landscape. The amount of over-aging, or the energy at this peak
strain, is due to the number of low energy states available to relax,
and therefore depends sensitively on the initial configuration.

This picture also helps us understand why over-aging occurs primarily
at low temperatures. At higher noise temperature, the low energy
states that are relaxed by small strains to produce over-aging are
rapidly depleted via thermal activation, meaning that the peak
over-aging is drastically reduced. Additionally, thermal aging results
in the relaxation of higher energy states during the strain cycle, and
these states are left with residual strain energy after the
cycle. Consequently, the final stress increases with noise
temperature, and the effective aging during the cycle is reduced.

\begin{figure}[tbp]
\begin{centering}
\includegraphics[width=8cm]{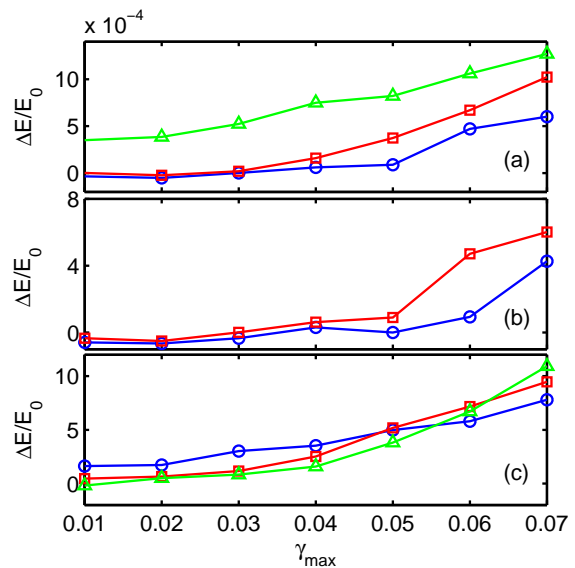}
	\caption{(Color online) Molecular dynamics results for the relative
	difference in energy versus strain after a single strain
	cycle. (a) $T=0.01$ (circle), 0.1 (square), and
	$0.2u_0/k_B$ (triangle). The quench time is $t_{qu}=750\tau_{LJ}$ and $\dot{\gamma}=5.3 \times
	10^{-5}\tau_{LJ}^{-1}$ for each sample. (b) $t_{qu}=750\tau_{LJ}$
	(square) and $7500\tau_{LJ}$ (circle). $T=0.01 u_0/k_B$
	and $\dot{\gamma}=5.3 \times 10^{-5}\tau_{LJ}^{-1}$. (c)
	$\dot{\gamma}=5.3 \times 10^{-6}\tau_{LJ}^{-1}$ (circle),
	$5.3 \times 10^{-5}\tau_{LJ}^{-1}$ (square), and $5.3 \times
	10^{-4}\tau_{LJ}^{-1}$ (triangle). $T=0.2u_0/k_B$, and $t_{qu}=750\tau_{LJ}$.}
	\label{fig:MD_strain}
	\end{centering}
\end{figure}

We compare these results with molecular dynamics simulations of the
model polymer glass under similar thermo-mechanical histories by
varying the maximum strain, the strain rate, and the temperature of
the glass. In order to investigate the effect of different initial
states, an equilibrated melt at $T=1.2u_0/k_B$ is quenched at
different rates to the final glassy temperature. The results are shown
in Fig.~\ref{fig:MD_strain} and are qualitatively similar to behaviour
found in the SGR model. Figure \ref{fig:MD_strain}(a) shows that
higher temperatures lead to increased rejuvenation,
Fig.~\ref{fig:MD_strain}(b) shows that initial states of higher energy
(faster quench) result in more rejuvenation, and
Fig.~\ref{fig:MD_strain}(c) shows increased rejuvenation at moderate
strains as well as decreased rejuvenation at large strains as the
strain rate is decreased. However, there is a significant difference
between the SGR and molecular dynamics results. We have not found
appreciable over-aging in our molecular dynamics simulations, even at
very low temperatures and fast cooling rates. This is in contrast to
recent molecular dynamics results of an atomistic polymer model under
a similar strain cycle ~\cite{Lyulin_PRL99}. The authors of this study
report over-aging of the simulated glass by the strain for very fast
quench rates.  However, this is likely because they did not compare to
a non-strained control sample but to the initial (just quenched)
energy before the strain cycle. We observe that even at very low
temperatures, our model exhibits significant aging directly after the
glass is quenched when quench rates are very high. It remains an open
question to what extent over-aging occurs in real polymer glasses
under realistic quench and loading conditions.

\begin{figure}[tbp]
\begin{centering}
\includegraphics[width=8cm]{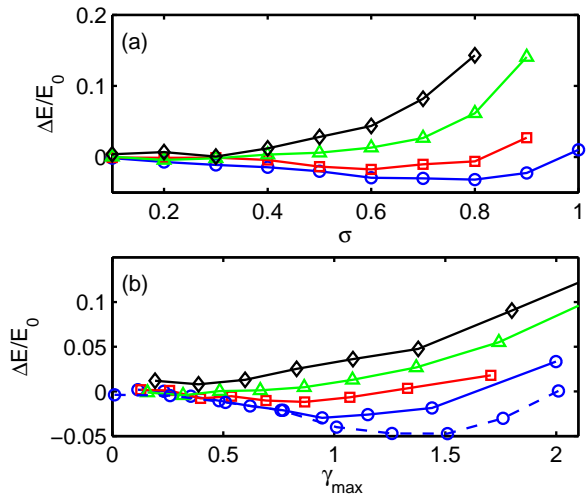}
	\caption{(Color online) Relative energy change due to a stress cycle
	applied for $t=10$. In (a) the energy difference is plotted
	versus the stress, and in (b) the same data is plotted versus
	the maximum strain in the stress cycle. $x=0.1$ (circles), 0.2 (squares), 0.4 (triangles),
	and 0.6 (diamonds). $x_l=2$ for each sample. The dashed
	line in (b) is the energy versus strain for an equivalent
	strain-controlled protocol at $x=0.1$. }
	\label{fig:stress}
	\end{centering}
\end{figure}

Stress-controlled loading protocols similarly show rejuvenation and over-aging
in the SGR model. In this case, a stress step function is applied for
a period of $t=10$ and then released. The change in energy due to
the stress cycle is plotted in Fig.~\ref{fig:stress}(a) for various
noise temperatures. At low noise temperature, there is over-aging at
low stress and rejuvenation at high stress. At high noise
temperature, there is a broad flat region for small stress, followed
by a steep increase in rejuvenation at high stress within the
nonlinear regime. In Fig.~\ref{fig:stress}(b), the same data is
plotted versus the maximum strain achieved during the stress cycle for
direct comparison with the strain-controlled data. The results plotted
in this way look very similar to Fig.~\ref{fig:strain}(a) for the
strain cycle. Over-aging occurs over a similar range of noise
temperature and strain, however, closer inspection shows that the
amount of over-aging is somewhat smaller for the stress-controlled
protocol. Finishing the cycle at zero stress rather than zero strain
means that low energy states that did not relax during the experiment
have residual strain energy and are rejuvenated.

\begin{figure}[tbp]
\begin{centering}
\includegraphics[width=8cm]{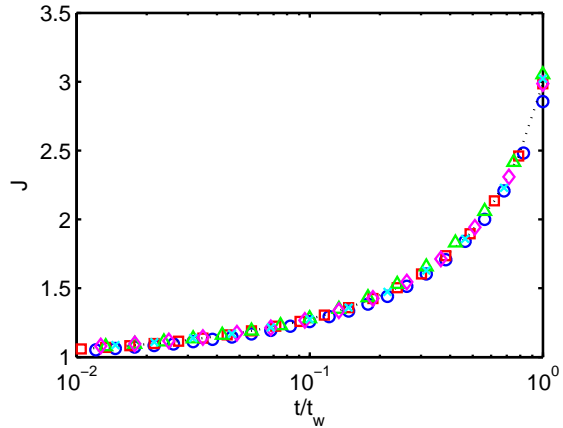}
	\caption{(Color online) Creep compliance of the SGR model quenched from
	$x_l=\infty$ (see text) and aged for $t_w=100$ (circles), 316
	(squares), 1000 (triangles), 3162 (diamonds), and 10000 ($\times$'s)
	at $x=0.5$ and $\sigma=0.75$ (within the nonlinear regime).}
	\label{fig:J_SGR}
	\end{centering}
\end{figure}

Our analysis so far has utilized changes in the energy of the system
to identify the occurrence of rejuvenation or over-aging.  However,
the notion of rejuvenation has its origin in an acceleration of the
dynamics under load
\cite{Struik,Cloitre_PRL85,Bonn_PRL89,Gacougnolle_Poly43,McKenna_JPhys15,Lequeux_PRL89}. As
described earlier, rejuvenation is commonly measured via a decrease in
the aging exponent $\mu$ with applied stress $\sigma$. Experimentally,
the aging exponent can be evaluated through creep experiments:
superposition of the creep compliance $J(t,t_w)$ of samples that have
aged for different wait times $t_w$ reveals the $t/t_w^\mu$ scaling
behaviour. In principle, this type of analysis can also be performed
in the SGR model, but is impeded by the fact that the creep compliance
exhibits a clear $t/t_w$ scaling only in the limit of $t_w \rightarrow
\infty$ \cite{Sollich_JR44}. In particular, for quenches from a finite
liquid temperature $x_l$, we find that the scaling regime is not
accessible within a reasonable simulation time. In the creep
experiments described below, we therefore initialize the trap
distribution from $\rho(E)$, where the scaling regime is more easily
accessible. This is equivalent to performing a quench from
$x_l=\infty$. Figure \ref{fig:J_SGR} shows representative compliance
curves for several different wait times under an applied load of
$\sigma=0.75$. Indeed, compliance curves obey the $t/t_w$ scaling
behaviour characteristic of trap models
\cite{Bouchaud_JPA}. Interestingly, we find this scaling behaviour to
be independent of the magnitude of the applied stress, even within the
nonlinear creep regime. Although the form of the scaling function is
stress dependent \cite{Sollich_JR44}, the decrease in the aging
exponent under large load that is observed in real structural glasses
does not occur in the SGR model.

\begin{figure}[tbp]
\begin{centering}
\includegraphics[width=8cm]{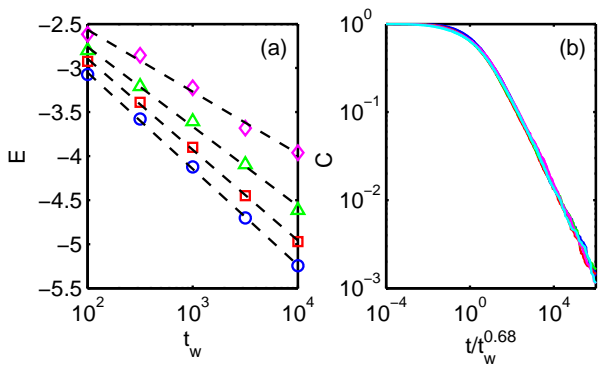}
	\caption{(Color online) (a) Energy as a function of wait time after a
	stress cycle of duration $t=t_w$. $\sigma=0.05$ (circles), 0.6
	(squares), 0.7 (triangles), 0.75 (diamonds). $x=0.5$ and
	$x_l=\infty$ for each sample. (b) Scaled correlation
	functions for $t_w=100$, 316, 1000, 3162, and 10000 (all
	overlapping) for $\sigma = 0.75$. Data collapse is obtained for
	$\mu=0.68$.}
	\label{fig:E_C_SGR}
	\end{centering}
\end{figure}

\begin{figure}[tbp]
\begin{centering}
\includegraphics[width=8cm]{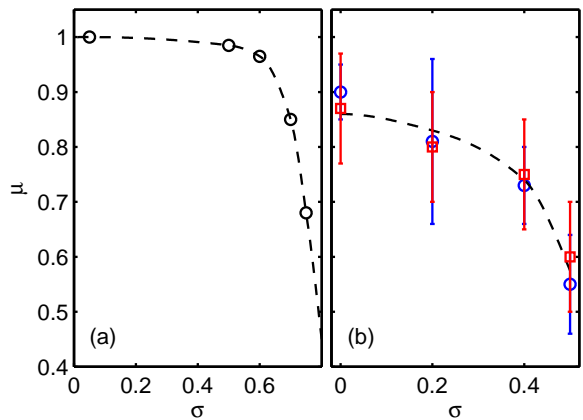}
	\caption{(Color online) (a) Aging exponent versus the stress in the SGR
	model, as determined from the correlation functions
	eq.~(\ref{corr-eq}) after a stress cycle of duration
	$t=t_w$. $x=0.5$ and $x_l=\infty$. (b) Aging exponent
	versus stress from molecular dynamics simulations of a model
	polymer glass using (circles) the creep compliance and
	(squares) the mean-squared displacement after stress release
	at $10t_w$. $T=0.2u_0/k_B$.}
	\label{fig:mu_stress}
	\end{centering}
\end{figure}

However, it is clear from the data in Fig.~\ref{fig:stress} that large
stress cycles cause rejuvenation of the energy of the
ensemble. To test the effect of wait time on the state of the aged
samples after creep, we release the stress after a creep experiment of
duration $t_w$ and measure the average energy of the unloaded system.
Figure \ref{fig:E_C_SGR}(a) plots the energy vs wait time for four
different stresses and shows that not only does the energy change
nonlinearly with the stress as anticipated from
Fig.~\ref{fig:stress}, but the slope of the energy versus wait time
curves also decreases for increasing stress. Highly stressed
samples appear to have aged less.

The simple relationship between the energy and the dynamics in the SGR
model implies a simultaneous change in the dynamics after the
stress is released. To see this, we compute the correlation
function
\begin{equation}
\label{corr-eq}
	C(t,t_w)=\int_0^\infty dE_b P(E_b,t,t_w)e^ {-\left( \Gamma_0 e^{-E_b/x}\right) t}
\end{equation}
which measures the probability that a particle in a trap at time $t_w$
has not hopped at time $t_w+t$ \cite{Bouchaud_JPA}. These functions
are shown in Fig.~\ref{fig:E_C_SGR}(b) for various wait times after a
nonlinear stress cycle. The shape of the correlation function is
typical of glasses, although missing the initial $\beta$-relaxation
decay. There is a plateau region at short times where there are very
few relaxation events, followed by a rapid decrease at $t\approx
t_w$. In contrast to the compliance curves, the scaling of correlation
functions with $t_w$ after the cycle is drastically changed by large
stresses. Figure \ref{fig:E_C_SGR}(b) shows that for $\sigma=0.75$, the
correlation functions no longer display full aging behavior, but we
find good data collapse if we rescale time with $t_w^\mu$, where
$\mu=0.68$. It appears that for stress controlled deformation at
finite temperature, the relaxation time spectrum of the mechanically
rejuvenated glass does indeed resemble a younger glass.

Figure \ref{fig:mu_stress}(a) presents the variation of aging
exponents with stress obtained from superposition of the correlation
functions at different wait times. At small stresses, the exponent is
unity, but decreases rapidly for nonlinear creep. For comparison with
the polymer model, we repeat the creep experiment using molecular
dynamics following a similar protocol. In MD, we extract aging
exponents from superposition of mean squared displacements (see
ref.~\cite{Warren_PRE07}) of particles after stress release. Figure
\ref{fig:mu_stress}(b) shows that these exponents (squares) similarly
decrease with increasing stress amplitude.  Remarkably, aging
exponents obtained by superposition of creep compliance curves
(circles) appear to be identical to those found through 
mean-squared displacements after the stress is released. This indicates
that the change in the aging exponent observed in nonlinear creep
experiments is due to the evolving configuration of the glass, rather
than the direct effect of stress on energy barriers, which would
cease when the stress is removed.

\begin{figure}[tbp]
\begin{centering}
\includegraphics[width=8cm]{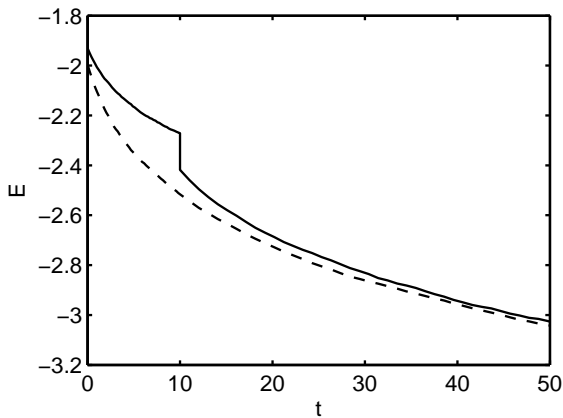}
	\caption{Energy during a stress cycle and the subsequent
	recovery (solid line) compared to an unstressed glass
	(dashed). Here $x=0.75$, $x_l=2$ and $\sigma=0.5$.}
	\label{fig:relaxation}
	\end{centering}
\end{figure}

It seems that the SGR model does produce rejuvenation in the
relaxation times, but is the mechanically rejuvenated glass actually
taken back in time by the application of stress? If this were the
case, we would expect the relaxation after stress to proceed exactly
like the younger, unstressed glass. In Fig.~\ref{fig:relaxation}, we
investigate the recovery of a glass after a stress cycle and compare 
to the natural aging trajectory of an unstressed glass. We see that
the aging progresses much more rapidly after the stress cycle, and
asymptotically approaches the energy relaxation of the unstressed
sample. This is the essence of the paradox pointed out by McKenna
\cite{McKenna_JPhys15}: although the stressed glass has an apparently
rejuvenated relaxation spectrum, the time it takes to reach
equilibrium is unchanged. In the SGR model, this has a simple
explanation. Every time a trap relaxes, it releases the strain it
was holding. Therefore, at constant (zero) stress, each relaxation
causes a decrease in the strain energy of the entire ensemble of
particles. This leads to a return to the unstrained aging trajectory
over a timescale similar to the decay of the correlation function. It
is conceivable that in real glasses, interactions between relaxing
elements similarly explain McKenna's observations.

\section{Conclusions}
We have performed stochastic simulations of the SGR model for
strain-controlled and stress-controlled loading cycles in the glassy
phase. Both types of loading induce changes to the mean energy of the
glass corresponding to regimes of mechanical rejuvenation and
over-aging. Within the SGR model, rejuvenation is the predominant
effect, while over-aging is only observed at low strains, low
temperatures, and high energy initial configurations. These results
are in qualitative agreement with molecular dynamics simulations of a
model polymer glass over a wide range of loading parameters.

In addition to changes in the energy of the glass, mechanical
rejuvenation is often observed as a decrease in the aging dynamics. To
this end, we have evaluated the aging exponent through the scaling of
the creep compliance with wait time, in direct analogy with
experiment. As previously remarked in ref.~\cite{Sollich_JR44}, the
creep compliance in the SGR model strictly obeys full aging ($\mu=1$),
even at very high load. In contrast, experiments
\cite{Struik,McKenna_Poly31,Cloitre_PRL85} and MD simulations
\cite{Warren_PRE07} show that the aging exponent decreases with stress
amplitude in the nonlinear creep regime. The physics of this
phenomenon does not seem to be captured by the SGR model; instead, the
dynamical effects of mechanical rejuvenation appear only after the
stress cycle. Aging exponents found from the $t_w$ scaling of 
correlation functions after unloading decrease with increased stress,
in qualitative agreement with MD simulations. Reconciling the stress
dependence of the aging exponent in the SGR model with experiment
would be a fruitful topic for further development of the model, but
may require a better understanding of the physics of mechanical
rejuvenation at the molecular level.

Finally, we have explored the relaxation dynamics after a rejuvenating
stress cycle. The SGR model exhibits the fundamental difficulty with
the ``rejuvenation'' hypothesis. While the relaxation times after a
stress cycle appear to be exactly identical to a younger glass, when
the stress is released, the aging trajectory gradually returns to the
undeformed path \cite{McKenna_JPhys15}. In the SGR model, this effect
is due to collective strain relaxation which occurs after each hopping
event. Identifying the effects of structural relaxations on the
rejuvenated state may similarly provide insight to this effect in real
glasses.

\end{document}